# Vacancy tuned thermoelectric properties and high spin filtering performance in graphene/silicene heterostructures

Zainab Gholami and Farhad Khoeini*

*Department of physics, University of Zanjan, P.O. Box 45195-313, Zanjan, Iran*

## Abstract

The main contribution of this paper is to study the spin caloritronic effects in defected graphene/silicene nanoribbon (GSNR) junctions. Each step-like GSNR is subjected to the ferromagnetic exchange and local external electric fields, and their responses are determined using the nonequilibrium Green's function (NEGF) approach. To further study the thermoelectric (TE) properties of the GSNRs, three defect arrangements of divacancies (DVs) are also considered for a larger system, and their responses are re-evaluated. The results demonstrate that the defected GSNRs with the DVs can provide an almost perfect thermal spin filtering effect (SFE), and spin switching. A negative differential thermoelectric resistance (NDTR) effect and high spin polarization efficiency (SPE) larger than 99.99% are obtained. The system with the DV defects can show a large spin-dependent Seebeck coefficient, equal to $S_s \sim 1.2$ mV/K, which is relatively large and acceptable. Appropriate thermal and electronic properties of the GSNRs can also be obtained by tuning up the DV orientation in the device region. Accordingly, the step-like GSNRs can be employed to produce high efficiency spin caloritronic devices with various features in practical applications.

## 1. Introduction

The next generation of electronic devices is often required to possess high performance, minimum waste heat, and low dimension in practical applications [1-4]. The spin caloritronics, as a new branch of spintronics, focuses on the relationship between the heat and spin transports in the material and attempts to pave the way for the development of nanostructure devices in the heat sensors, waste heat recyclers, and future device technologies [5-7]. The spin-dependent

Seebeck effect (SDSE), as an essential feature of the spin caloritronics, aims to produce spin-polarized currents by using the temperature gradient. This can efficiently help us to obtain a pure spin current with no charge companions. The spin equivalent of the classical Seebeck effect was first observed by Uchida *et al.* [8] in the ferromagnetic metals. Numerous studies were then performed by other researchers on quasi-one or two-dimensional (2D) materials such as graphene (GE), silicene (SE), $MoS_2$, and so on [9-12]. Although nanoribbon materials with zigzag edges have inherent antiferromagnetic properties, they can be magnetized by using an external magnetic field. This can provide us to understand well SDSE and thermal spin transport attributes of such types of materials (e.g., spin-dependent Seebeck diode (SSD) effect, thermal SFE, and thermal spin-giant-magnetoresistance (GMR) effect [13-15]).

GE and SE are now introduced as the famous family of low-dimensional materials [16-18] with unusual features for the next generation of device applications [19,20]. They consist of a honeycomb structure with promising electronic properties such as high carrier mobility [21-24]. GE and SE nanostructures are now identified as materials with adequate Seebeck coefficient and widespread spin caloritronics applications [21,23-25]. Quasi-one-dimensional structures, theoretically, show satisfactory thermoelectric performance compared to 2D structures [26]. This may be attributed to the fact that cutting a 2D material in to a nanoribbon can decrease the thermal conductivity and lead to better thermoelectric properties. A GE-based thermoelectric device was experimentally introduced by Sierra *et al.* [27], to produce a spin voltage via two ferromagnetic thermal sources with variant temperatures. However, GE is not often identified as an efficient thermoelectric material, because it shows a low figure of merit (FOM) [28]. In contrast, SE has superior thermoelectric properties than GE and can drastically enhance the Seebeck coefficient, due to its nonzero band gap [24,29]. According to previous studies, the zigzag SE nanoribbons (ZSNRs) can suitably provide a significant spin filter efficiency [30], and a perfect SDSE [9]. The thermoelectric performance of nanostructures can also be enhanced by various ways, such as hybridization [31,32], defects [33-35], and doping [36,37]. As shown in the previous studies [4,38,39], the hybrid nanostructures theoretically show superior thermoelectric properties than the similar materials with the single nanostructures. Several studies have already been performed to investigate various properties of the GSNRs in which GE and SE have vertically or laterally bonded together [40-43]. The results showed that the combination of GE with other 2D materials could lead to a hybrid structure with various properties and applications [44-47]. Nevertheless, there are little data about the electronic behavior, spin transport, and thermoelectric attributes of the GSNRs [38,46], and further research is still required in this area.

Several studies have been conducted to investigate the effect of structural defects and pattern geometry on the thermal and thermoelectric phenomena of nanoribbons [10,48-50]. Experimental results clearly show that the thermal conductivity of GE with a patterned vacancy (nanomesh) structure is significantly lower than that observed by the pristine GE [51]. In addition, nanoribbons with periodic edge defects, called saw tooth-like (ST) [52], can also improve the spin thermoelectric properties, identified by the large values of spin-dependent Seebeck coefficient and FOM [53]. Some designs have been used to connect the GE flake to two armchair GE nanoribbons (AGNRs), and to generate spin-dependent currents [54,55]. The width of the nanoribbon, the coupling between the leads and the ribbon, and the topology were also identified as the main factors affecting the electrical and thermal properties of the nanoribbon. Sonvane *et al*. showed that the thermal transport of a GNR could significantly be affected by changing its length or width [56]. Some unusual thermal and electrical transport properties have been observed in the asymmetric shapes of nanoribbons [57-60]. All these observations may be related to the symmetric or asymmetric connection of the device to the leads and quantum interference [61]. The thermal and thermoelectric properties of the GNRs have been studied by Rojo *et al*. [62], and Tan *et al*. [10], for the vertical and parallel step-like junctions, respectively. However, there is still little information about the thermoelectric phenomena of hybrid junctions with different geometry and structural defects. This is especially true for the parallel step-like GSNRs, inspiring us to explore the thermal transport attributes in this type of hybrid nanostructures.

In this study, the spin caloritronic effects in the parallel step-like graphene/silicene nanoribbon (GSNR) junctions, including the divacancy, are numerically studied using the tight-binding (TB) and NEGF approaches. Three GSNRs with different length-to-width ratios are first selected and then subjected to the ferromagnetic exchange and local external electric fields. There are many experimental studies available in the literature, which have focused on developing various nanostructures [40,63-69], and measuring their spin thermoelectric properties [70,71]. However, this paper numerically investigates the thermoelectric properties of parallel step-like GSNRs because: (1) the nanoribbons can satisfactorily show thermoelectric performance compared to their corresponding 2D structures [72] (2) the step-like nanostructures can provide some unusual thermal and electrical transport properties for the asymmetric shape of nanoribbons [62] (3) the hybrid type of nanostructures can often provide superior thermoelectric properties compared to the similar materials with the single nanostructures [32], and (4) the defected nanostructures can provide different thermoelectric properties of the nanoribbons. They can also provide more realistic simulations for the actual experimental tests [73]. Nonetheless, the calculation of the spin thermoelectric properties of the considered GSNRs using experimental methods is beyond the scope of the present research. Although there may be a difference between the results obtained by this research with those given by the actual experimental tests, the results of this study can suitably pave the way for the development of standard experimental methods and thermoelectric property assessment of the GSNRs. It is concluded that a nearly full spin filtering effect with spin-dependent localized transmission peaks near the chemical potential μ = 0 can be obtained. Moreover, larger spin-dependent Seebeck coefficient and spin polarization efficiency are identified compared to those given by [74-76]. To further study the thermoelectric properties of the GSNRs, three defect arrangements with different orientations are also considered for the larger case study, and their responses are re-evaluated. The results of the present study are summarized in the subsequent sections.

## 2. Device structure and computational details

Three different parallel step-like GSNR junctions are introduced using semi-infinite metallic AGNRs with different widths of the left and right leads, and a ZSNR in the central region [43,77]. Each left and right lead is generated by duplicating two different unit cells along the transport direction. As shown in figure 1, each system contains $2N_{AG}$ = 46, 70, and 94 carbon atoms in each unit cell in the left and $N_{AG}$ -1 in the right region. $N_{AS}$ = 6, 10, and 14 number of armchair silicon-atom chains are also considered for the central regions. The changes in the central region width are dependent on the left contact size. These step-like GSNR junctions are named 23GE-6SE-11GE, 35GE-10SE-17GE, and 47GE-14SE-23GE, respectively. To further study the divacancy effects on the electronic and thermoelectric properties of the selected devices, three different defect orientations are considered in the central scattering region. In this study, the pentagon-octagon-pentagon (5-8-5) defect type [78,79] is selected for the defective system. The DVs are also assumed to be duplicated with the periodicity of $l$ = 4 [80]. Despite a DV in the nanostructure, an equal number of A- and B-sublattice sites are assumed in all cases (see figure 1). This can provide net magnetic moment per unit cell i.e., $m = (N_A - N_B) \mu_B = 0$ [81].

To obtain the spin caloritronic characteristics of the selected nanostructures, their electron transmissions are first determined using the NEGF method [82] and Landauer-Bütticker transport formula [83]. Because there is a weak coupling among the electrons and phonons, the electron-phonon interactions are ignored in this study [84-86]. Thus, electronic transports can appropriately be assumed ballistic. The spin-flip scattering is also neglected. Hence, the spin-up and spin-down electron transports can singly be investigated. This is valid because the spin diffusion length in SE is in the order of several micrometers [87]. As shown in figure 1, the system is divided into three various segments, including the left and right electrodes, and the central region. The generalized Hamiltonian for the system consists of several submatrices as follows:

$$H_{\text{T}} = H_{\text{L}} + H_{\text{R}} + H_{\text{C}} + H_{\text{CL}} + H_{\text{CR}}, \tag{1}$$

where $H_{\text{L}}$, $H_{\text{R}}$, and $H_{\text{C}}$ are the Hamiltonian matrices of the isolated left electrode, right electrode, and the central region, respectively. These submatrices can be determined by using Eqs. (2) and (3). In Eq. (4), $H_{\text{CL}}$ and $H_{\text{CR}}$ are also defined by the hopping between the central region and the left and right electrodes, respectively. The Hamiltonian matrices are all determined using the TB method [88-90].

$$H_{\text{R(L)}} = -t_{\text{R(L)}} \sum_{<i,j>,\alpha} c_{i\alpha}^{\dagger} c_{j\alpha} + eE_{YG} \sum_{i,\alpha} Y_{iG} c_{i\alpha}^{\dagger} c_{i\alpha} + \text{H.c.}, \tag{2}$$

$$H_{\text{C}} = -t_{\text{C}} \sum_{<i,j>,\alpha} c_{i\alpha}^{\dagger} c_{j\alpha} + i \frac{\lambda_{so}}{3\sqrt{3}} \sum_{\ll i,j \gg,\alpha,\beta} v_{ij} c_{i\alpha}^{\dagger} (\sigma_z)_{\alpha\beta} c_{j\beta} + M_Z \sum_{i,\alpha} c_{i\alpha}^{\dagger} \sigma_z c_{i\alpha} + \tag{3}$$

$$+elE_Z \sum_{i,\alpha} \xi_i c_{i\alpha}^{\dagger} c_{i\alpha} + eE_{YS} \sum_{i,\alpha} Y_{iS} c_{i\alpha}^{\dagger} c_{i\alpha} + \text{H.c.},$$

$$H_{\text{CR(CL)}} = -t_{\text{CR(CL)}} \sum_{<i,j>,\alpha} c_{i\alpha}^{\dagger} c_{j\alpha} + \text{H.c.}, \tag{4}$$

in which $t_{\text{R(L)}}$ and $t_{\text{C}}$ are the hopping energies for the nearest-neighbor interactions in the right (left) lead, and the central region, respectively. In this study, 2.66 and 1.60 eV values are assumed for $t_{\text{R(L)}}$ and $t_{\text{C}}$, respectively [88,90]. $c_{i\alpha}^{\dagger}$ ($c_{i\alpha}$) denotes the fermion creation (annihilation) operator at site $i$. $\alpha$ also is the spin index. $<i, j>$ and $\ll i, j \gg$ show the nearest-neighbor and second-nearest-neighbor atoms, respectively. The rest of parameters used in Eqs. (2) − (4) are defined as follows: $\lambda_{so}$ is the effective spin-orbit interaction and is assumed equal to 3.9 meV and zero for the SE and GE nanoribbons, respectively [89]. $\sigma$ is the Pauli matrix. $v_{ij}$ is a parameter equal to −1 (+1) for the clockwise (counterclockwise) hopping index about the Z-axis. $\xi_i = +1, -1$ is the valley index for the sublattices $A$ and $B$, respectively. $E_Z$ is a perpendicular electric field that can yield a voltage difference equal to $2elE_Z$ in which e is the electron charge and $l$ is the half distance of the two sublattices. $M_Z$ is the exchange field induced by the proximity effect of ferromagnetic material [91]. $E_{YG}$ and $E_{YS}$ are the inhomogeneous transverse electric fields along $y$-direction, and applied to the leads and central region, respectively. $Y_{iG}$ and $Y_{iS}$ denote the $y$-coordinate of the atom $i$ in the SE and GE nanoribbons. In this study, the electric and ferromagnetic exchange fields are assumed to be 0.082 V/Å and 0.162 eV, respectively, and perpendicularly applied to the central region of the hybrid GSNRs. The effects of inhomogeneous transverse electric fields $E_{YS}$ = 0.917 V/Å and $E_{YG}$ = 0.680 V/Å are also considered. $t_{\text{CR}}$ ($t_{\text{CL}}$) denotes the hopping energy between the central region and right (left) lead, respectively. In this study, Harrison's scaling law [92,93] is used to compute the contact hopping energies for the GE and SE at the interface. At the interface, the bond length between carbon and silicon atoms is 1.89 Å, and its corresponding hopping energy is calculated as $t_{CR(CL)}$=1.88 eV. It is worth mentioning that this value needs to be in the same order of $t_{R(L)}$ and $t_C$ [4]. The spin-dependent transmission function, $T_\alpha$, is used to determine the thermoelectric properties of the nanostructures, and can be obtained by Eq. (5) and using NEGF formalism [94]:

$$T_\alpha(E) = \text{Tr}\big[\Gamma_{L,\alpha}(E) G_{C,\alpha}(E) \Gamma_{R,\alpha}(E) G_{C,\alpha}^\dagger(E)\big],$$ (5)

where $\Gamma_{R(L),\alpha} = i\big(\sum_{R(L),\alpha} - \sum_{R(L),\alpha}^\dagger\big)$ is the interaction between the right (left) lead with the scattering region, in which $\Sigma_{R(L),\alpha}$ is the right (left) self-energy [94]. $\Sigma_{R(L),\alpha}$ includes the effect of two semi-infinite AGNRs on the scattering region. $G_{C,\alpha}(E)$ also denotes the retarded spin Green's function of the scattering region, and can be expressed as follows [94]:

$$G_{C,\alpha} = \big[(E + i0^+)I - H_C - \sum_{L,\alpha} - \sum_{R,\alpha}\big]^{-1},$$ (6)

The temperatures in the left and right leads are set to $T_L$ and $T_R$, respectively. Hence, the temperature difference can be obtained by $\Delta T = T_L - T_R$. Based on the Landauer-Büttiker formula, the spin-dependent current can be obtained by [95]:

$$I_\alpha = \frac{e}{h} \int_{-\infty}^{+\infty} T_\alpha(E)[f_L(E, T_L) - f_R(E, T_R)]\, dE,$$ (7)

where e and h are the electron charge and the Plank constant, respectively. $f_{R(L)}$ and $T_{R(L)}$ represent the average Fermi-Dirac distribution and the temperature for the right (left) lead, respectively. $T_\alpha$ is also the spin-dependent transmission. The net spin current is defined as $I_s = I_{up} - I_{dn}$ and calculated by Eq. (7). Considering a linear response regime (i.e., $\Delta T \cong 0$), the spin-dependent Seebeck coefficient can then be computed for each spin channel as follows [96]:

$$S_\alpha = -\frac{1}{eT}\left(\frac{L_{1,\alpha}}{L_{0,\alpha}}\right).$$ (8)

where $L_{n,\alpha}$ is the intermediate function and defined as [95]

$$L_{n,\alpha}(\mu, T) = -\frac{1}{h}\int(E - \mu)^n \frac{\partial f(E, \mu, T)}{\partial E} T_\alpha(E) dE,$$ (9)

where $\mu$ is the chemical potential. The spin and charge Seebeck coefficients can then be computed by $S_s = S_{up} - S_{dn}$ and $S_c = (S_{up} + S_{dn})/2$, respectively [96]. The electrical conductance is also defined as

$$\mathrm{G}_\alpha = -e^2 \big( L_{0,\alpha} \big). \tag{10}$$

where $G_c = G_{up} + G_{dn}$ and $G_s = \mid G_{up} - G_{dn} \mid$ is the charge and spin thermal conductance, respectively [97]. The spin-dependent electron thermal conductance is obtained as [98]

$$\kappa_{\mathrm{e},\alpha} = \frac{1}{T} \left( L_{2,\alpha} - \frac{L_{1,\alpha}^2}{L_{0,\alpha}} \right). \tag{11}$$

$$Z_{\mathrm{e},c(s)} T = S_{c(s)}^2 \mathrm{G}_{c(s)} T / \kappa_{\mathrm{e}}. \tag{12}$$

where $Z_{\mathrm{e}}T$ is the electrical FOM and is defined as the upper limit of $ZT = S^2 GT/(\kappa_{\mathrm{e}} + \kappa_{\mathrm{ph}})$ value, in which $\kappa_{\mathrm{ph}}$ and $\kappa_{\mathrm{e}}$ are the phonon's and electron thermal conductance, respectively. The maximum value of $ZT$ occurs when $\kappa_{\mathrm{ph}}$ is very smaller than $\kappa_{\mathrm{e}}$ [99,100].

### 3.1. Analyses and results

To numerically investigate the performance of the step-like GSNRs, three different length-to-width ratios (β) of the central region, i.e., β≅ 0.9, 1.0, and 1.1, are selected for the considered devices (see figure 1). To produce a step-like nanoribbon, it is assumed that the right contact in all cases has a half-width from the left one. Each device is subjected to perpendicular electric ($E_Z$) and ferromagnetic exchange ($M_Z$) fields in the central region. It is noted that both local $E_Z$ and $M_Z$ fields [101] can be produced in the laboratory environment. The effect of inhomogeneous transverse electric fields, i.e., $E_{YS}$ and $E_{YG}$ are also included for each system, respectively [102]. The electric current is also calculated by applying a difference between the temperature of the left, $T_L$, and the right, $T_R$, electrodes with no back gate voltage, and the spin currents are only determined in the presence of the temperature gradient.

In each system, a temperature gradient is generated between the two electrodes using $\Delta T > 0$. This can cause a nonzero value of $f_L$ - $f_R$, and produce a spin-dependent current depending on $T_{R(L)}$ and $\Delta T$. In addition, it creates changes in the spin-up and spin-down currents in the GSNRs at different step-like junctions. The thermal SFE is studied for the considered hybrid step-like GSNRs with threshold temperature ($T_{\mathrm{th}}$). The variations of the spin-dependent current $I_{up}$ ($I_{dn}$) versus $T_R$ with $\Delta T = 10$, 20, and 40 K are shown in figures 2a to 2c, representing small to large step-like GSNRs. As shown in these figures, when $T_R$ increases, $I_{up}$ reaches its maximum value for the high temperatures, whereas $I_{dn}$ value is almost zero and remains unchanged for the whole temperature range. Moreover, the spin-up current is positive ($I_{up} > 0$) for almost $T_R > 100$ K, while the spin-down current $I_{dn} \cong 0$ throughout. The results show that the considered

GSNRs can well show an insulating response with no charge or spin current for nearly $T_R < 100$ K. For example, the threshold temperature value $T_{th} = 50$ K occurs at $\Delta T = 40$ K for the 23GE-6SE-11GE, whereas this value for the 35GE-10SE-17GE and the 47GE-14SE-23GE cases is similarly equal to 30 K. In addition, $T_{th}$ marginally decreases by increasing the GSNR size. This is dependent on where the peak of spin-up transmission occurs as the system size changes (see figure 4). Thus, $T_{th}$ value can be controlled by changing the system size. This clearly illustrates that the SFE [15] is accompanied by the spin switching effect in these cases. As shown in figures 2b and 2c, the SFE is stronger when the device size increases. The peak values of $I_{up}$ are almost obtained as 4.2, 10.24, and 10.83 nA for the 23GE-6SE-11GE, 35GE-10SE-17GE, and 47GE-14SE-23GE case studies, respectively (see figures 2a to 2f). Thus, the strength of the current varies when the device size is changed. However, the NDTR happens for the higher temperatures and $\Delta T$. Accordingly, the hybrid step-like GSNRs can be utilized as a thermal spin device with multiple different characteristics. Figures 2d to 2f also show the changes of $I_{up}$ and $I_{dn}$ against $\Delta T$ for $T_R = 200, 250,$ and 350 K, respectively. These figures demonstrate that $I_{up}$ increases as $T_R$ and $\Delta T$ increase, while $I_{dn}$ keeps nearly zero for various $\Delta T$ values. This again confirms that the thermal SFE has been generated. A thermal rectification behavior is also observed for the lower $T_R$ and larger size of devices (see inset of figure 2f).

Figure 3 shows the variation of the spin polarization efficiency SPE (= $(|I_{up}|-|I_{dn}|)/ (|I_{up}|+|I_{dn}|) \times 100\%$) versus $T_R$ and $\Delta T$ for the studied systems. The results illustrate that the considered step-like GSNR junctions can suitably provide a thermal spin current with high SPE. The SPE of the devices is often dependent on the temperature sets and, most notably, on their sizes. Therefore, it is not appropriate to study the SPE below the 'on-off' temperature for the spin-up currents, where the spin-up and spin-down channels are closed. The results show that a nearly 100% SPE can be achieved for a broad range of $T_R$ and $\Delta T$ values. This is especially true for the 47GE-14SE-23GE configuration (see figures 3e and 3f). Figure 3 also shows that SPE experiences sudden changes for a limited range of temperatures in some cases. This is due to the $I_{up}$ reverse sign and the mutual competition among $I_{up}$ and $I_{dn}$ within this range of temperatures (See insets in figure 3). The SPE values larger than 99.80% and 99.99% can also be observed for the second and third case studies for the different values of $\Delta T$ at $T_R = 200$K, respectively, whereas a nearly 93.0% value is obtained for the first case. This evidence confirms that the GNR and SNR-based devices often show less SPE values than their corresponding studied counterparts [74,75,103]. The results of numerical analyses clearly show that the response of a hybrid nanostructure strongly depends on the values selected for the external electric and ferromagnetic exchange fields. The value of the fields in the present study has been chosen in such a way that we obtain the SFE and maximize

its efficiency. The results demonstrate that the selected magnitudes can provide high-efficiency value up to 99.9% in some case studies. It is noted that such a high SFE value has previously been obtained and reported by other researchers (e.g.,[104-107]).

The changes of the spin-polarized transmission coefficients ($T_{up}$ and $T_{dn}$) against energy ($E - E_F$) and spin-dependent electrical conductance ($G_{up}$ and $G_{dn}$) versus the chemical potential ($\mu$) for the three studied cases are displayed in figure 4. In these plots, $T_{up}$ ($T_{dn}$) and $G_{up}$ ($G_{dn}$) are shown by a filled area and dotted line, respectively, and the Fermi level is set to zero. As shown in figure 4, the transmission peak values almost locally occur around the Fermi level i.e., the peak value of $T_{up}$ almost occurs at [-0.16eV, -0.06eV], [-0.001eV, -0.07eV], and [-0.17eV, -0.07eV] intervals, for the 23GE-6SE-11GE, 35GE-10SE-17GE, and 47GE-14SE-23GE configurations, respectively. Nonetheless, they do not include a strong peak value for $T_{dn}$ around the Fermi level. This is particularly true for the second and third configurations. This confirms that the spin-up governs the transport properties with $I_{up} > 0$, and can generate an almost perfect thermal SFE. The spin gap along with the narrow transmission peaks near the chemical potential $\mu = 0$ eV can lead to a larger spin-dependent Seebeck coefficient compared to the values provided by the GNRs and SNRs [108,109]. Figure 4 clearly shows that $G_{dn}$ is nearly zero in the spin-down channel for some ranges of $\mu$ values, whereas $G_{up}$ is conductive in the spin-up channel. As a result, the proposed nanostructure device can provide a fully spin-polarized current.

Figure 5 shows the variation of $S_{up}$, $S_{dn}$, $S_s$, and $S_c$ versus the chemical potential ($\mu$) at $T_R = 300$ K. As shown in Figure 5, the response of $S_{up}$ ($S_{dn}$) is not very identical around $\mu = 0$ eV for various sizes of nanoribbons. For each configuration, there are also some points at a given chemical potential value in which $S_\alpha$ (or thermal spin current) is equal to zero for one spin channel and is not zero for the other one; because the electron and hole currents are neutralized by one another. Hence, a perfect spin-polarized current can be obtained by the thermal gradient. Based on the results, the peak values of $S_s$ around $\mu = 0$ eV are obtained as 0.404, 0.491, and 0.547 mV/K for the 23GE-6SE-11GE, 35GE-10SE-17GE, and 47GE-14SE-23GE case studies, respectively [76]. The results also show that $S_s$ can be increased as the device size is increased. This may be attributed to the fact that the spin gap gets larger. The peak value of $|S_s|$ occurs in the large chemical potential region and almost around $\mu = \pm 0.4$ eV, while the location of these peak values almost remained unchanged when the device size increases. The maximum values of $|S_s|$ for the 23GE-6SE-11GE, 35GE-10SE-17GE, and 47GE-14SE-23GE cases are obtained as 0.826, 1.150, and 1.114 mV/K, respectively. These values are very larger than the values reported in other researches (e.g., $S_s \sim 300$ μV/K in [97,98], $S_s \sim 550$ μV/K in [76], and $S_s \sim$ -90 µV/K in [108]). The color of $S_s$ varies in accordance with the Seebeck polarization, $P_s$ (= ($|S_{up}|$ - $|S_{dn}|$)

/ ($|S_{up}|$ + $|S_{dn}|$)) in figure 5. As shown in figure 5, $S_s$ color varies from orange to black, representing the contribution of the spin-down and spin-up electrons in $S_s$, respectively. The charge and spin thermoelectric efficiencies are also assessed by calculating the electrical FOM ($Z_{e,c}T$ and $Z_{e,s}T$) for each case study and different µ values at room temperature (See insets of figure 5). The results show that the thermoelectric efficiency of a given device can significantly be improved as the size changes. The maximum values of the charge (spin) thermoelectric efficiencies occur about 15.06 (101.4), 13.4 (49.2), and 4.5 (9.2) from the large to small devices, respectively. It is realized that the third nanostructure case study (i.e., 47GE-14SE-23GE) can provide the largest charge and spin thermoelectric efficiency and enhance $Z_eT$. This enhancement is very larger than that reported in [110]. The obtained results illustrate that the engineering defects in step-like GSNRs can provide a quite acceptable spin thermoelectric efficiency. It is noted that $Z_eT$ is always larger than $ZT$. Hence, high value for $Z_eT$ is nonsense in reality. Low $Z_eT$ values are important because they allow us to exclude unpromising materials.

## 3.2. The effect of DV orientation on spin caloritronics

In order to achieve complete spin-dependent thermoelectric properties of the step-like GSNR junctions, a set of DVs with three different orientations in the 47GE-14SE-23GE (i.e., DV1, DV2, and DV3) is created and are analytically assessed [78,111] (see figure 1c). The variation of $I_{up}$ and $I_{dn}$ against $T_R$ for DV1, DV2, and DV3 cases and $\Delta T = 40$ K are displayed in figure 6. $I_{up}$ and $I_{dn}$ values for no vacancy (NV) defect configuration are also plotted for the sake of comparison. As shown in the inset of figure 6a, the spin and charge transporting channels are almost closed for the

NV and DV3 cases for the entire temperature ranges. In contrast, the other cases (i.e., DV1 and DV2) respond as an insulator for $T_R$< 40 K. In DV1 and DV2 cases, $I_{dn}$ almost remained unchanged and equal to zero for the entire range of temperatures, while $I_{up}$ is significantly increased as $T_R$ exceeds 40 K. This confirms that the SFE is along with the spin switching effect in these cases, and occur only due to the temperature gradient with no back gate voltage. The

NDTR phenomenon occurs for the latter two cases. For each case, $I_{up}$ absolute value also increases by $T_R$, while their flowing directions are opposite at a given $T_{th}$. Figure 6b also presents the variation of $I_{up}$ ($I_{dn}$) against $\Delta T$ for $T_R = 350$

K. As shown in figure 6b, $I_{up}$ almost linearly increases for the DV1, whereas it is decreased for the DV2.

The variations of SPE against $T_R$ and $\Delta T$ are also separately plotted for DV1 and DV2 case studies in figure 7. $\Delta T =$

40 K and $T_R = 350$ K are assumed for these plots, respectively. An SPE value larger than 99.6% can be obtained for

DV2 case and a broad range of $T_R$ and $\Delta T$ values. This shows a nearly 0.5% SPE value difference between DV1 and DV2 [78]. In the DV2 case, the maximum SPE value occurs for high $T_R$ and low $\Delta T$ values. This is fully compatible with the results reported in figure 6. It is worth mentioning that a sudden drop can be seen in figure 7 for a limited range of $T_R$ values. However, this is negligible and is not essential in practical applications.

Figure 8 shows the variations of $T_{up}$ ($T_{dn}$) against $E - E_F$ and $G_{up}$ ($G_{dn}$) versus $\mu$ for DV1 and DV2 cases. The plot indicates that the variation of DV orientation can change the values of $T_{up}$ ($T_{dn}$) and $G_{up}$ ($G_{dn}$), while keeping the spin-up electron as the dominant carrier in both cases. Comparing DV1 and DV2 cases shows that $T_{up}$ is remarkably enhanced when the divancancy orientation is changed to DV1. Although the peak values occur around $\mu = 0$ eV in both cases, there are two peaks for DV1 case at [-0.17eV, -0.07eV], whereas two peak values for DV2 case are in both sides of this point, i.e., the first peak at [-0.17eV, -0.11eV] and the second one at [0.07eV, 0.11eV] interval. This indicates that the current sign for the DV2 case is negative and changes to positive with the increase of $T_R$ and NDTR. In DV1, the maximum values of $T_{up}$ and $G_{up}$ are almost obtained 30% and 90% larger than the stronger value in the DV2 (see figure 8). As a result, $I_{up}$ in the DV2 case is negative and weaker. Nevertheless, the values of $T_{dn}$ and $G_{dn}$ are nearly zero in both cases and confirm the SFE. Accordingly, the spin-dependent transmission coefficient, the direction, and intensity of the current can be engineered using the DV orientation.

Figure 9 depicts the variation of $S_{up}$, $S_{dn}$, $S_s$, and $S_c$ versus $\mu$ for the studied cases at $T = 300$ K. By comparing $S_S$ values obtained with the DV1 and DV2 configurations, we find that the change of DV orientation may vary $S_S$. As shown in figure 9, the maximum value of $S_S$ is almost equal to 1.114 mV/K for the DV1 case, whereas, this value is 0.748 mV/K for the DV2 case. In the figure, $S_s$ color varies from orange to black, representing the contribution of the spin-down and spin-up electrons in $S_s$, respectively. The charge and spin thermoelectric efficiency are also assessed by calculating the electrical FOM ($Z_{e,c}T$ and $Z_{e,s}T$) for each case study, and different $\mu$ values at room temperature (See insets in figure 9). The maximum values of $Z_{e,c}T$ and $Z_{e,s}T$ for the DV1 case are relatively 14% and 99% higher than those obtained by the DV2 case. This provides evidence that the thermoelectric efficiency of the case studies is controlled and enhanced by the DV.

## 4. Conclusions

In this research, the thermal SFE, spin-dependent electronic, and thermoelectric properties of the defected step-like GSNR junctions have numerically been studied. Each configuration was then subjected to the ferromagnetic exchange and local external electric fields. The divacancy effects were considered in the studied cases using the pentagon-octagon pentagon (5-8-5) type model. Higher amounts of spin-up current were also determined for high-temperature values, while the spin-down current is remained unchanged and equal to zero for all temperature values, when the right lead temperature ($T_R$) increases. Hence, a satisfactory SFE and a spin polarization up to 99.99% were obtained in the studied GSNR junctions. The magnitude of the electric and ferromagnetic exchange fields is optimally selected to achieve the spin filtering effect with maximum efficiency. However, the SFE and current strength become more robust when the device size increases. This response behavior with localized transmission peaks could be achieved around the Fermi level if the divacancy was well oriented in the nanostructure. This provides evidence that the spin-dependent transmission coefficient, the direction, and intensity of the current can be engineered using the DV orientation. The NV and DV3 case studies showed insulator behavior, whereas the two other cases showed spin filtering behavior but with opposite spin current directions. It is realized that the DV1 case study (i.e., 47GE-14SE-23GE) can provide the strongest SFE and largest spin thermoelectric efficiency and enhance $Z_eT$. Some interesting transport properties such as the spin switching effect, rectifying behavior of the devices, and NDTR in thermal charge current were also observed. The maximum values of $|S_s|$ and the charge (spin) thermoelectric efficiencies for the 47GE-14SE-23GE case are obtained as 1.114 mV/K and 15.06 (101.4), respectively. These values are satisfactory and may be comparable to the values reported in other studies [98,108,110]. In general, the considered hybrid GSNRs can be used for thermoelectric applications using different system temperature sets.

*Corresponding author: Farhad Khoeini (khoeini@znu.ac.ir)

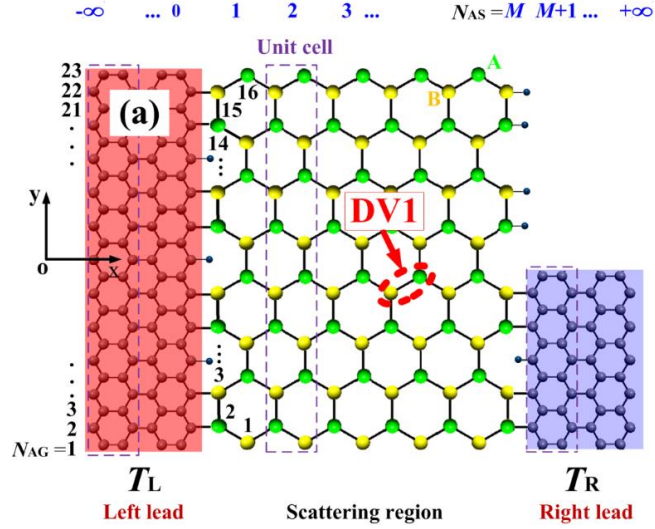

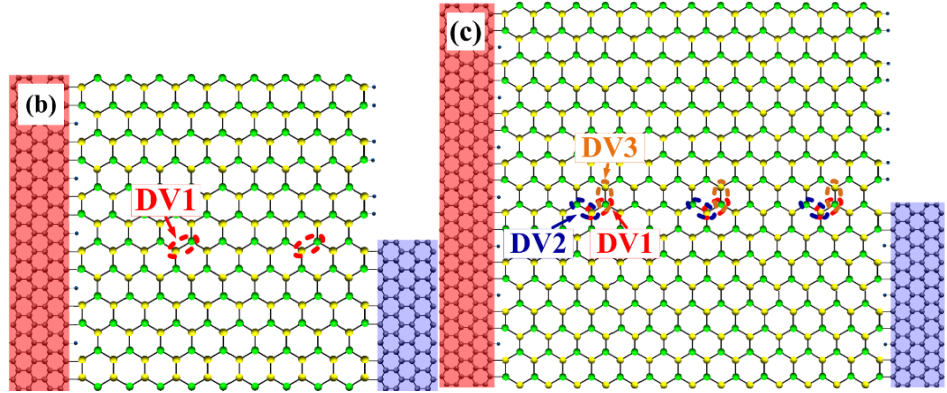

**Figure 1**. Schematic view of the defected parallel step-like GSNR junctions for (a) the 23GE-6SE-11GE, (b) the 35GE-10SE-17GE, and (c) the 47GE-14SE-23GE case studies. Each unit cell of the left lead has $2N_{AG}$ atoms. The central region is composed of about $N_{AS}$ unit cells. Different DV orientations are also represented in the 47GE-14SE-23GE.

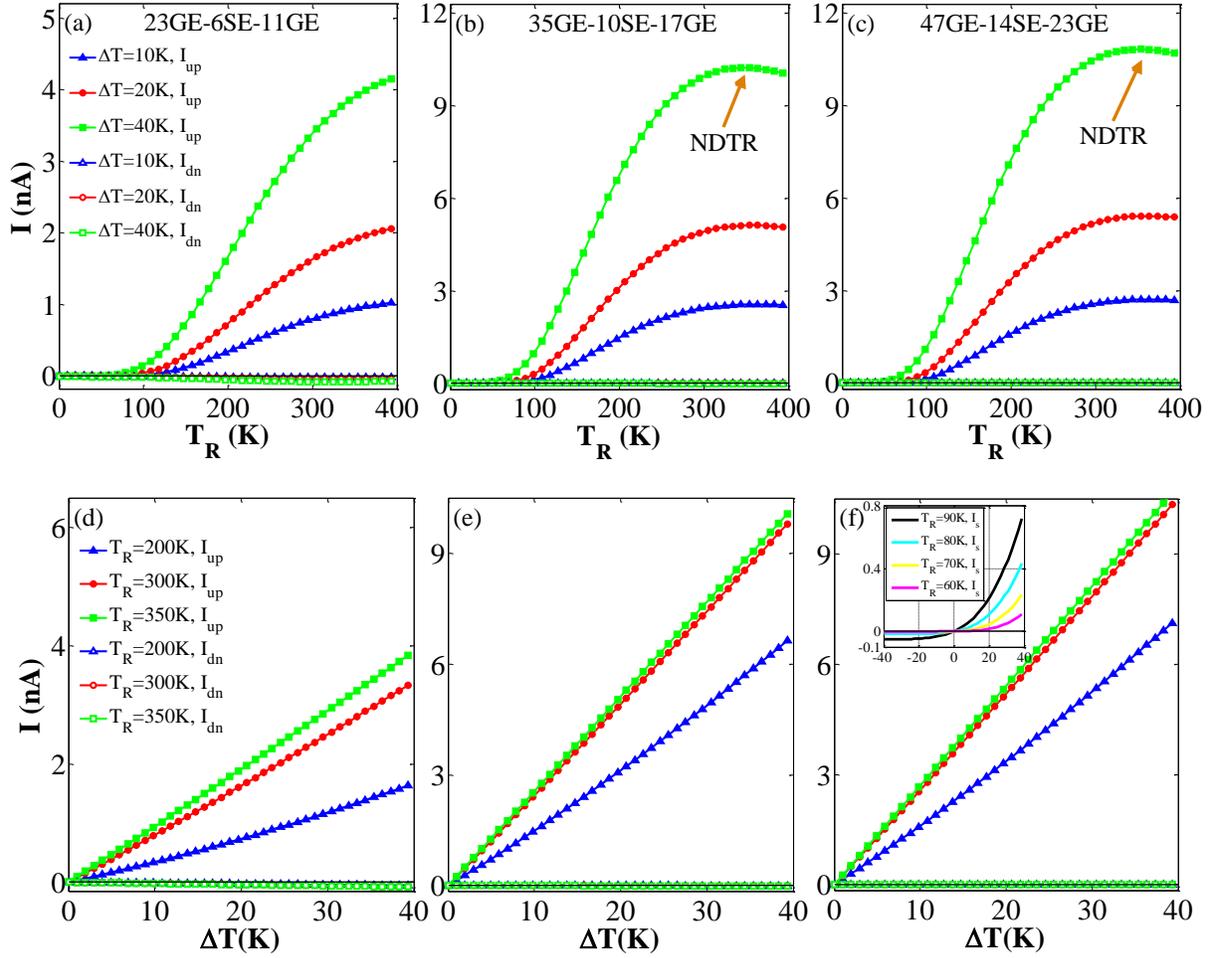

**Figure 2**. Panels (a), (b), and (c) show the variation of the spin-polarized currents ($I_{up}$ and $I_{dn}$) against $T_R$ for $\Delta T$= 10, 20, and 40 K; Panels (d), (e), and (f) show $I_{up}$ and $I_{dn}$ against $\Delta T$ for $T_R$ = 200, 250, and 350 K for the 23GE-6SE-11GE, 35GE-10SE-17GE, and 47GE-14SE-23GE configurations, respectively.

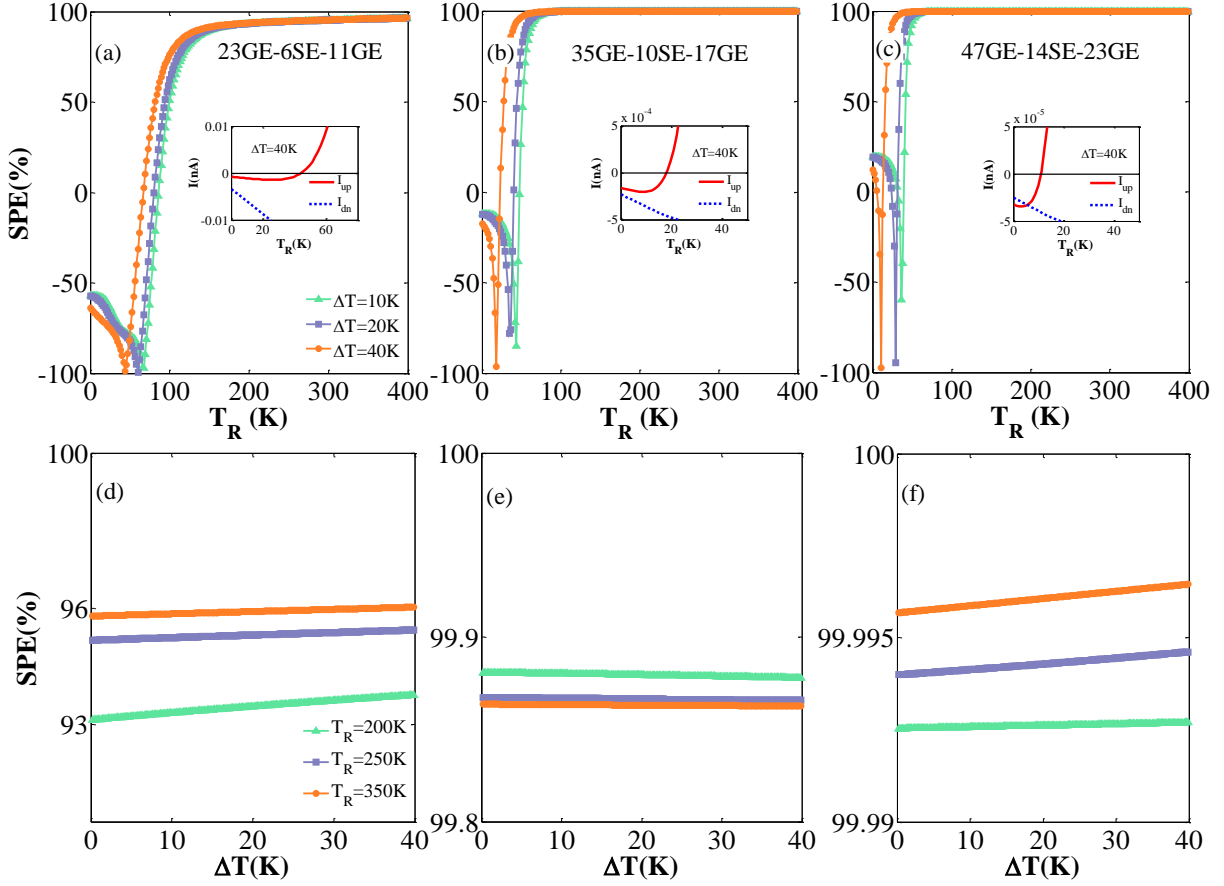

**Figure 3.** Panels (a), (b), and (c) show the variation of the SPE against $T_R$ for $\Delta T = 10$, 20, and 40 K; Panels (d), (e), and (f) show the SPE variation against $\Delta T$ for $T_R = 200$, 250, and 350 K for the 23GE-6SE-11GE, 35GE-10SE-17GE, and 47GE-14SE-23GE configurations, respectively. The insets show the $I_{up}$ reverse sign and the mutual competition among $I_{up}$ and $I_{dn}$ for $T_R < 100$ K and $\Delta T = 40$ K.

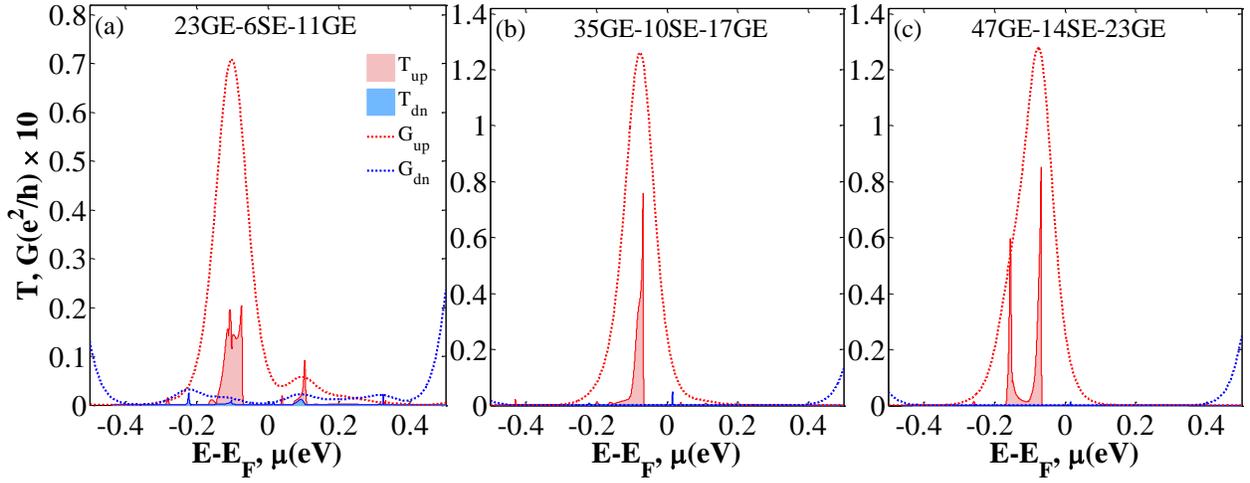

**Figure 4**. Panels (a), (b), and (c) show the variation of the spin-dependent transmission spectra ($T_{up}$ and $T_{dn}$) against $E\text{-}E_F$ (filled area) and dotted lines show the variation of the spin-dependent electrical conductance ($G_{up}$ and $G_{dn}$) versus $\mu$ for the 23GE-6SE-11GE, 35GE-10SE-17GE, and 47GE-14SE-23GE configurations, respectively, at $T =$ 300 K.

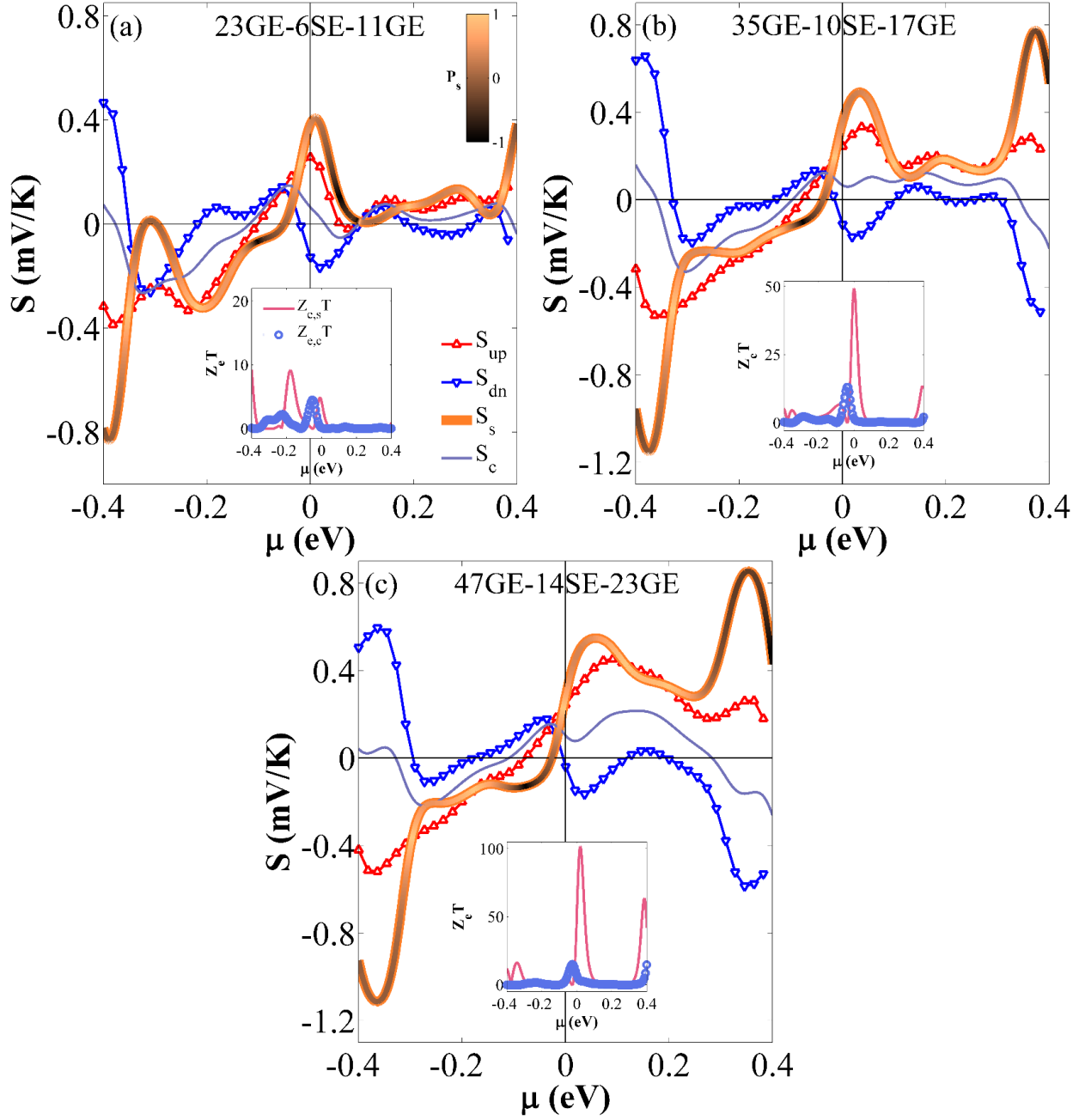

**Figure 5.** The $S_{up}$, $S_{dn}$, $S_s$, and $S_c$ against $\mu$ at room temperature for (a), (b), and (c) 23GE-6SE-11GE, 35GE-10SE-17GE, and 47GE-14SE-23GE configurations, respectively. $S_S$ color varies based on Seebeck polarization ($P_s$) value. The insets show the $Z_{e,s}T$ and $Z_{e,c}T$ as a function of $\mu$.

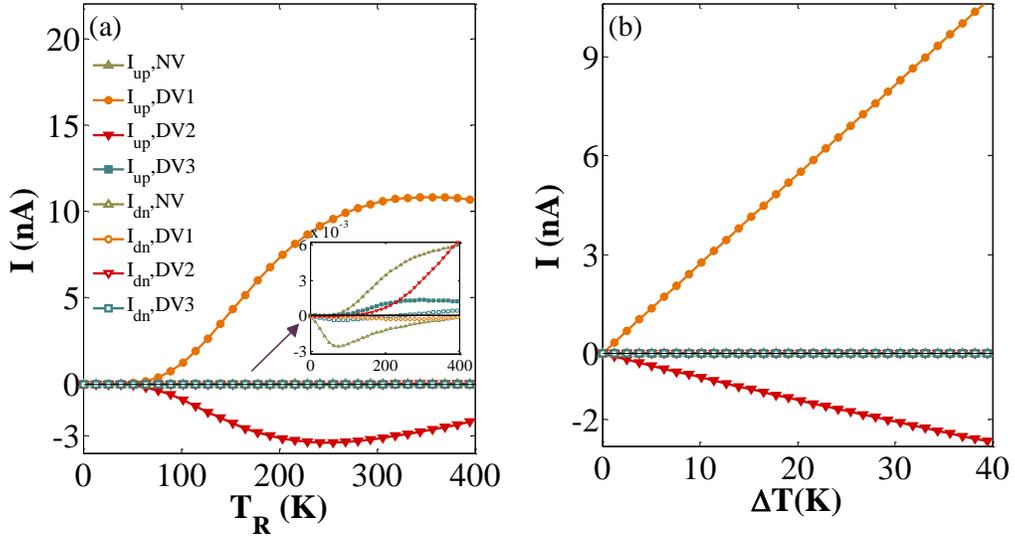

**Figure 6.** (a) $I_{up}$ and $I_{dn}$ versus $T_R$ for $\Delta T$ = 40 K; (b) $I_{up}$ and $I_{dn}$ versus $\Delta T$ for $T_R$ = 350 K and NV, DV1, DV2, and DV3 configurations, respectively.

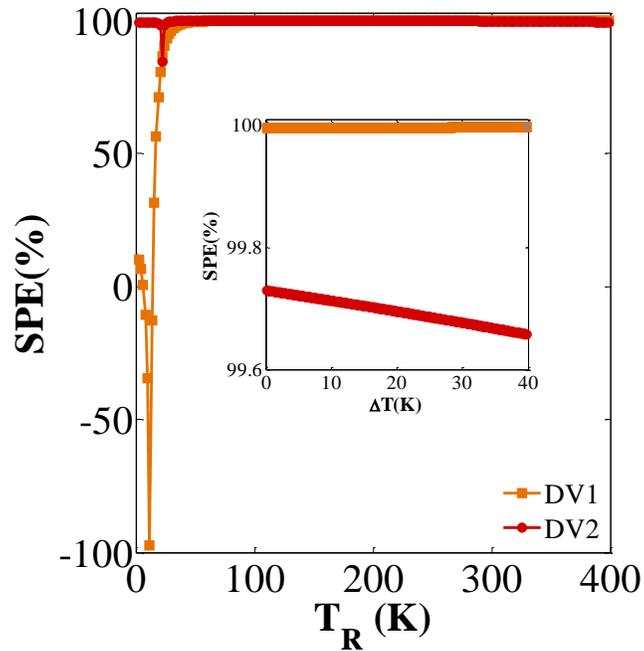

**Figure 7.** The SPE versus $T_R$ for $\Delta T$ = 40 K; The inset shows SPE versus $\Delta T$ for $T_R$ = 350 K; for DV1 and DV2 configurations, respectively.

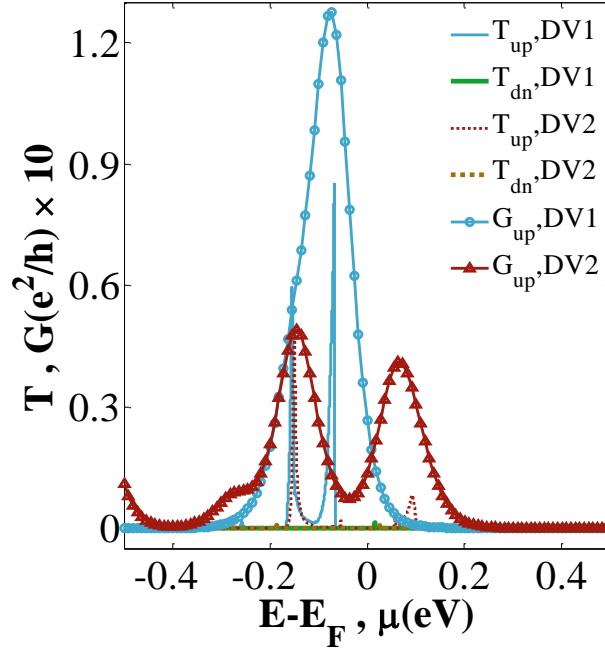

**Figure 8.** $T_{up}$ and $T_{dn}$ channels versus $E$-$E_F$, and $G_{up}$ versus $\mu$ for the DV1 and DV2 configurations, at $T$ = 300 K, respectively.

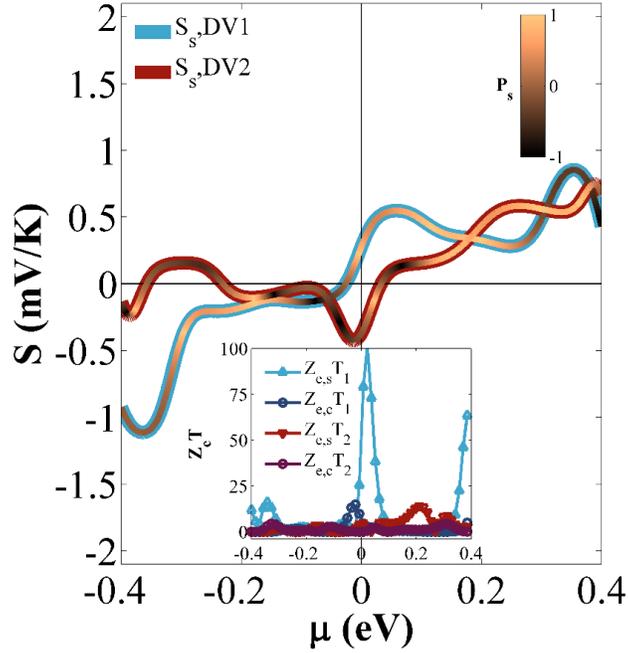

**Figure 9.** $S_{up}$, $S_{dn}$, $S_s$, and $S_c$ against $\mu$ at room temperature for DV1 and DV2 cases, respectively. $S_S$ color varies based on Seebeck polarization ($P_s$) value. The inset shows $Z_{e,s}T$ and $Z_{e,c}T$ as a function of $\mu$.